\documentclass[10pt,conference]{IEEEtran}
\usepackage{xcolor}
\usepackage{tabularx}
\usepackage{flushend}

\usepackage{pgf}
\usepackage{tikz}
\usetikzlibrary{arrows,automata}
\usepackage{verbatim}

\usepackage{arydshln}

\usepackage{cite}
\usepackage{url}
\usepackage{graphicx}
\usepackage{framed}

\graphicspath{{./diagrams/}}

\newcommand{\se}{software engineering}

\begin{document}
\title{How does Software Change?}
%\title{A Theory of Software Change}
 \author{\IEEEauthorblockN{Ayushi Rastogi}
 \IEEEauthorblockA{University of Groningen, the Netherlands\\
 a.rastogi@rug.nl}
 \and
 \IEEEauthorblockN{Georgios Gousios}
 \IEEEauthorblockA{Facebook\\
 gousiosg@fb.com}
 }
\maketitle
\begin{abstract}
% Context
Software evolves with changes to its codebase over time.
Internally, software changes in response to decisions to include some code change into the codebase and discard others.  
Explaining the mechanism of software evolution, 
this paper presents \emph{a theory of software change}.
Our theory is grounded in multiple evidence sources 
(e.g., GitHub documentation and relevant scientific literature)
relating to the pull-based development model in GitHub.
The resulting theory \emph{explains} the influence of project-related core concepts (e.g., people and governance) as well as its ecosystem on the decision of software change. 
\end{abstract}
\IEEEpeerreviewmaketitle

\section{Introduction}

A distinct characteristic of software systems is that they continuously change;
as Lehman's first law of software evolution states: ``a [\ldots] system must 
be continually adapted or it becomes progressively less satisfactory"~\cite{lehman1979understanding}.
Software evolves to adapt to new functional or non-functional requirements, 
or due to unforeseen circumstances that necessitate its adaptation (e.g., fixing
security problems).
Traditionally, software evolution was analyzed and understood in the context 
of enterprise software projects, who undergo development in discrete phases,
such as initial development, servicing, phasing out, etc. (the staging model\cite{rajlich2000staged}).
The emergence of the open source development model in the late 1990's,
along with the tooling that facilitated it, made incremental,
piece-wise software change (and its analysis) a first class citizen in 
software evolution research.

% Describe software evolution
Within modern software teams, software evolves by integrating small changes, 
%commonly referred to as \emph{patches} and \emph{pull requests}
(hereafter referred to as \emph{unit changes}), 
%into its codebase.
which originate in changing user requirements or predictive/corrective
maintenance cycles.
Unit changes usually undergo some form of explicit inspection~\cite{bacchelli2013expectations}
by the software maintenance team and as a whole (both code changes and
inspection results) the unit change is recorded in software's version
control system.
We refer to this process of changes to software code based on review decision as 
\emph{modern software change} or, more generally \emph{software change}.
Ultimately, modern software evolves by assimilating decisions to update its 
codebase over time, while evidence that led to those decisions is maintained.
 
% Relevance of software change in general
Evolution through integration of unit changes is a prevalent development model 
in open source development~\cite{beller2014modern} 
(ranging from the Linux kernel development process to development on platforms like GitHub~\cite{gousiosExploratoryStudyPullbased2014b})
and large industrial systems today~\cite{bacchelli2013expectations, sadowski2018modern, KRHDG19}.
This standardization has led to software development processes able to scale projects 
to thousands of developers working in a distributed manner on multiple locations world wide.
Moreover, standardization on evolution through unit changes 
also facilitated the introduction and wide spread use of technologies such as continuous delivery\cite{yuDeterminantsPullbasedDevelopment2016a} 
and platforms, such as GitHub, that further facilitate software change
by automating tedious tasks.

% Relevance of unit change in SE research
The relevance of unit changes in understanding software evolution attracted widespread investigation in software engineering research.
From exploring the characteristics of unit changes~\cite{gousiosDatasetPullbasedDevelopment2014c} and  
factors influencing them~\cite{ramWhatMakesCode2018}, 
to factors that influence decisions~\cite{soaresAcceptanceFactorsPull2015b} on whether to
accept or reject them~\cite{soaresAcceptanceFactorsPull2015b},
and going even as far as online community dynamics~\cite{tsayInfluenceSocialTechnical2014b} 
and personality traits~\cite{iyerEffectsPersonalityTraits2019},
researchers have investigated individual aspects of unit changes and the various processes
around them in depth. %to death
With much empirical evidence explaining parts of the process of software 
change (including replication studies), what is missing is a framework that 
will enable us to understand the combined effects of individual factors
on how software changes through unit changes.

% Objective and scope
Our goal with this study is to synthesize the available empirical evidence 
by building a theory explaining the mechanism of modern software change.
Our theory is derived from the evidence on collaboratively developed software 
projects built using pull-based development model on GitHub.
The resulting \emph{process theory} is \emph{grounded in empirical data} 
explaining the decision process of evaluating unit changes and the 
factors that affect it in terms of abstract concepts and relations. 

% Method
We build our theory incrementally, based upon multiple sources of evidence.
First, we ground our theory on practical experience, i.e., GitHub documentation
%\footnote{https://docs.github.com/en} 
and analysis of selected software projects.
%(e.g., rails\footnote{https://github.com/rails/rails} and facebook-ios-sdk\footnote{https://github.com/facebook/facebook-ios-sdk})).
Next, we revise the concepts and categories derived from practical experience 
based on the relevant scientific literature in software engineering.
In this phase, we also add relations to the abstract concepts and categories.

% Results
Our theory surfaces six core concepts that must interact in order for software to change:
\emph{available unit changes} define an interdependent set of items that need to be evaluated;
\emph{existing codebase} is the result of the application of past unit changes and defines the context under which
software change occurs;
\emph{people} propose unit changes, evaluate them, and ultimately make the change decisions;
\emph{(project) governance} defines the context and processes followed when evaluating unit changes;
\emph{tools and technology} facilitate unit changes by automating processes;
and the \emph{ecosystem} defines the technical and social norms under which unit changes are being evaluated.
Moreover, our theory surfaces a set of relationships that define the effects of the interactions on 
%\ar{each of the entities and on} 
the decision to accept a unit change.
%A framework of concepts and categories influencing software change is shown in Figure~\ref{fig:initialTheoryFramework}, followed by an initial theory in Figure~\ref{fig:initialTheory}.
%The final theory of software change is shown in Figure~\ref{fig:finalTheory}.

% Consolidate contributions
Our theory of software change serves as:
\begin{enumerate}
\item a \emph{reference document} for understanding the mechanism of software change;
\item a \emph{comprehension} of the existing scientific literature presented as simple relations;
\item a guide for researchers to position their work in the broader \emph{frame of reference} offered by our theory; and
\item as direct inspiration for future work, through a set of \emph{derived propositions}.
\end{enumerate}

The complete paper is organized into nine sections.
Section~\ref{sec:background} gives a brief introduction to the necessary background.
Section~\ref{sec:design} presents our design to build the theory.
The next two sections present our concepts, initial theory (in Section~\ref{sec:initialTheory}) and final theory of software change (in Section~\ref{sec:finalTheory}).
% and its evaluation (in Section~\ref{sec:evaluation}).
We discuss our theory and its implications in Section~\ref{sec:discussion}, followed by related work (in Section~\ref{sec:relatedWork}) and threats to validity (in Section~\ref{sec:threatsToValidity}).
Finally, Section~\ref{sec:conclusion} presents a summary of our theory and potential directions for future research.

\section{Background}
\label{sec:background}
% Outline
In this section, we introduce the definition of theory, its goals, and ways to evaluate it.
Next, we give a brief history on the ways software change from code inspection to modern code review practices 
(for details see systematic mapping in~\cite{wang2019evolution}).
Finally, we describe a widely used collaborative development model - pull based development model~\cite{gousios2016work}
which is also the focus of this theory.

\subsection{An Introduction to Theory}
% Introduction to theory
%% What is theory?
There are many definitions of theory, with most definitions sharing the following three characteristics~\cite{gregor2006nature}: 
 \begin{enumerate}
\item attempt to \emph{generalize} local observations and data into more abstract and universal knowledge; 
\item represent \emph{causality} (cause and effect); and 
\item aim to \emph{explain or predict} a phenomenon. 
\end{enumerate}

Generally, a theory is build to meet one or more of the following four goals~\cite{gregor2006nature}: 
\begin{enumerate}
\item simply describe the studied phenomenon;
\item explain the how, why, and when of the topic; 
\item explain what has already happened but also to predict what will happen next; and
\item prescribe how to act based on predictions.
\end{enumerate}

%% How to evaluate a theory?
Finally, once a theory is build, its theoretical quality is evaluated on 
(1) consistency,
(2) correctness,
(3) comprehensiveness, and
(4) precision~\cite{gregor2006nature}.

% Traditional ways of software change
\subsection{History of Software Change}
In 1970s, changes to codebase were preceded by \emph{code inspection} -- 
``a structured and formal process in which authors and reviewers sit together in a room, inspecting the proposed changes for defect''~\cite{fagan1976design}. 
Reviews have changed since then in practices as well as purpose.

\emph{Modern code review} is a ``lightweight process conducted iteratively with small changes discussed quickly and frequently by few(-er) reviewers synchronously or asynchronously in time''~\cite{rigby2013convergent}.
The purpose of code review has evolved from merely finding defects to group problem solving but also to maintain team awareness, improve code quality, assess high-level design and more (e.g., see~\cite{bacchelli2013expectations}).

Some examples of tools used for modern code review in software companies are CodeFlow by Microsoft~\cite{bacchelli2013expectations}, Phabricator by Facebook\footnote{https://techcrunch.com/2011/08/07/oh-what-noble-scribe-hath-penned-these-words/}, and CRITIQUE by Google~\cite{sadowski2018modern}).
Some other general-purpose code review tools are Code Collaborator\footnote{https://smartbear.com/product/collaborator/overview/} and Crucible.\footnote{https://www.atlassian.com/software/crucible}
Its equivalent in open source are Gerrit\footnote{https://www.gerritcodereview.com/} (CRITIQUE's open source equivalent by Google), ReviewBoard\footnote{https://www.reviewboard.org/} by VMware, and pull based development model in GitHub.\footnote{https://docs.github.com/en/github/collaborating-with-issues-and-pull-requests/about-collaborative-development-models}

% Characteristics of the theory
\begin{table*}
\caption{Characteristics of our theory}
\label{tab:characteristics}
\begin{tabular}{l l}
Type of theory & \emph{evolutionary process theory}~\cite{ralph2018toward} explaining how software changes  \\
View & \emph{Positivist} (verifiable by observations or experience) as well as \emph{Interpretivist} (understand from the views of those who live it) \\
Breadth of area & \emph{Mid-range theory} offering a moderate abstraction with limited scope leading to testable hypotheses \\
%Causality & \emph{Probabilistic causal analysis} since there exists effects of many extraneous influences \\
Method & \emph{Constructivist grounded theory} based on close and careful analysis of data\\
\end{tabular}
\end{table*}

% Pull-based development model
\subsection{Pull-based Development Model}
In pull-based development model,\footnote{https://docs.github.com/en/github/collaborating-with-issues-and-pull-requests/about-collaborative-development-models} a contributor wanting to make changes to a project forks an existing git repository.
Forking creates a copy of the existing repository at the local machine of the contributor,
although a copy can be made within the project itself (referred to as branch).
While a fork can serve many purposes,\footnote{https://help.github.com/en/enterprise/2.13/user/articles/fork-a-repo} 
a pull request includes the changes that are candidate for inclusion into the primary line of development and is available for review.

Contributor makes changes to their fork and send it out as a pull request.
A pull request include changes to code in one or more files.
The pull request appears in the list of pull requests for the project in question, visible to anyone who can see the project.
Next, contingent on its adoption, the pull request is reviewed and decision is made to merge it in the codebase as it is, with some modifications, or not merge it.
When a pull request is merged, code changes in the fork are committed into the codebase.
Currently, this model is widely used in open source (e.g., GitHub, Bitbucket, and Gitorious) as well as software companies (e.g., Microsoft, Facebook, and Google).

\section{Design}
\label{sec:design}
% What type of theory is being build?
Our approach to build theory is inspired by several available resources.
The general procedure to build theory is inspired by the guides on building theories in software engineering~\cite{sjoberg2008building} and building theory from multiple evidences~\cite{shull2008building}. 
For specific guidelines to build process theory grounded in qualitative data, we refer to Ralph's guidelines for process theories~\cite{ralph2018toward} and Charmaz's constructivist grounded theory approach~\cite{charmaz2014constructing}.
Generally, we take inspiration from existing theories in \se{} research (such as~\cite{herbsleb2006collaboration, franca2012towards,bjarnason2016theory, baltes2018towards, storey2019towards}), details on which are available in Section~\ref{sec:relatedWork} on related work. 
	
\subsection{Theory formation}
% Steps in theory formation
%% How theories are build?
Theories in \se{} are inspired from (e.g.,~\cite{storey2019towards}) or are adaptations (e.g.,~\cite{herbsleb2006collaboration}) of 
other theories in the related disciplines of Social Science, Organizational Psychology and Distributed Systems.
Our theory of software change, instead, is build from scratch; given the specificity of the subject to \se{}.
Although later, we show support for our theory in the form of empirical evidence instantiating the propositions derived from our theory.
A summary on the characteristics of our theory is shown in Table~\ref{tab:characteristics}. 

%% How we choose to build theory? Overview
Since building theory is an iterative process of proposing, testing, and modifying theory~\cite{sjoberg2008building}, 
we build our theory in two phases.
First, we ground our theory in the empirical data derived from \emph{practical experience}.
Later, we revise the resulting theory based on the relevant \emph{scientific literature} in software engineering. 

%%What resources we use in phase I?
We use three information sources as proxy for practical experience on 
collaboratively developed software projects build using the pull-based development model on GitHub.
We analyze all information on the GitHub website,\footnote{https://github.com/} its official documentation,\footnote{https://docs.github.com/en} and available survey reports\footnote{https://opensourcesurvey.org/2017/} as primary source.
We do not include white papers, blogs, and webcasts on GitHub since they are likely to offer opinions, not facts.
Next, we select secondary sources of information such as \emph{open source guides}\footnote{https://opensource.guide/} and \emph{GHTorrent}~\cite{gousiosGHTorrentGitHubData2012} for breadth.
Finally, we include four software projects and their documentation including README files, CONTRIBUTION guidelines, and Code of Conduct documents as proxy for project vision, development guidelines, and community guidelines respectively.
We analyzed the \emph{rails} project as representative of a large popular open source software and \emph{facebook-ios-sdk} as its company equivalent.
We also analyzed two relatively less popular projects written in Java\footnote{https://github.com/SquareSquash/java} and Python,\footnote{https://github.com/emitter-io/python} respectively. 

%% What resources we use in phase II?
We expanded on the evidence from practical experience by adding insights from the relevant scientific literature in \se{}.
We searched the six most popular search engines -- ACM, Springer, IEEE, ScienceDirect, Web of Science, and Google Scholar using the query: (``pull request'' OR ``pull based'') AND (``software'' OR ``development'') AND ``github''.
In this query, terms ``software'' and ``development'' ensure that the resulting papers relate to software engineering.
Other terms - ``pull request'', ``pull based'', and ``github'' refer to relevant aspects of the theory and the scope of this work.
We identified 1206 papers in total, out of which 250+ papers where found relevant based on title and abstract. 
For each paper, we asked ourselves the following three questions for selection. Does this paper potentially identify ...:
\begin{enumerate}
\item concept(s) influencing software change?
\item influence of concept(s) on software change?
\item relationships among concepts?
\end{enumerate}

The resulting selection of papers reached theoretical saturation in expressing some concepts and relations while other concepts and relations where still relatively under developed.
To bridge this gap, we collected additional papers to build up the less developed concepts and relations.

% How do we analyze the qualitative data collected in phase I?
To build grounded theory from the evidence collected in the above, we followed Charmaz's constructivist grounded theory~\cite{charmaz2014constructing} approach.
Constructivist grounded theory is the most recent and one of the three most widely followed schools of grounded theory~\cite{stol2016grounded}.
Constructivist approach divides the analysis process in three main phases: 
(1) initial coding,
(2) focused coding and categorization, and 
(3) theory building. 
During \emph{initial coding}, we examined the data word-by-word, line-by-line, as well as incident-by-incident without introducing our assumptions~\cite{stol2016grounded}.
Next, during \emph{focused coding}, we categorized the codes identified during initial coding into core categories~\cite{stol2016grounded}.
Finally, during \emph{theoretical coding} we specified the relations between categories to generate a cohesive theory~\cite{stol2016grounded}. 

% How did we analyze?
All coding for building the grounded theory was performed by the first author. 
Later, the second author collaborated with the first author to derive strong interpretations via mutual agreement~\cite{silva2020engagement}. 
Further, to ensure that the abstractions can be traced to its source raw data, we are sharing the list of resources used and the intermediate results as a part of replication package.~\footnote{https://figshare.com/s/16f491d0e2cfb487f2cf}
% To do

%% Evaluation
\subsection{Theoretical quality}
We evaluate the goodness of our theory on the six aspects described by Sj{\o}berg et al.~\cite{sjoberg2008building}: 
\begin{itemize}
\item Is it possible to empirically refute? (\emph{Testability})
\item Do empirical studies confirm its validity? (\emph{Empirical support})
\item To what degree do the theory account for and predict all known observations? (\emph{Explanatory power}) 
\item Is the theory economically constructed with minimum concepts and propositions? (\emph{Parsimony})
\item What is the breadth of scope and theory independence of setting? (\emph{Generality})
\item What is the use of the theory? (\emph{Utility})
\end{itemize}

Additionally, we discuss the \emph{credibility, originality}, and \emph{usefulness} of our theory as criteria relevant for evaluating constructivist grounded theory~\cite{stol2016grounded}. 

\begin{figure*}
\includegraphics[width=\textwidth]{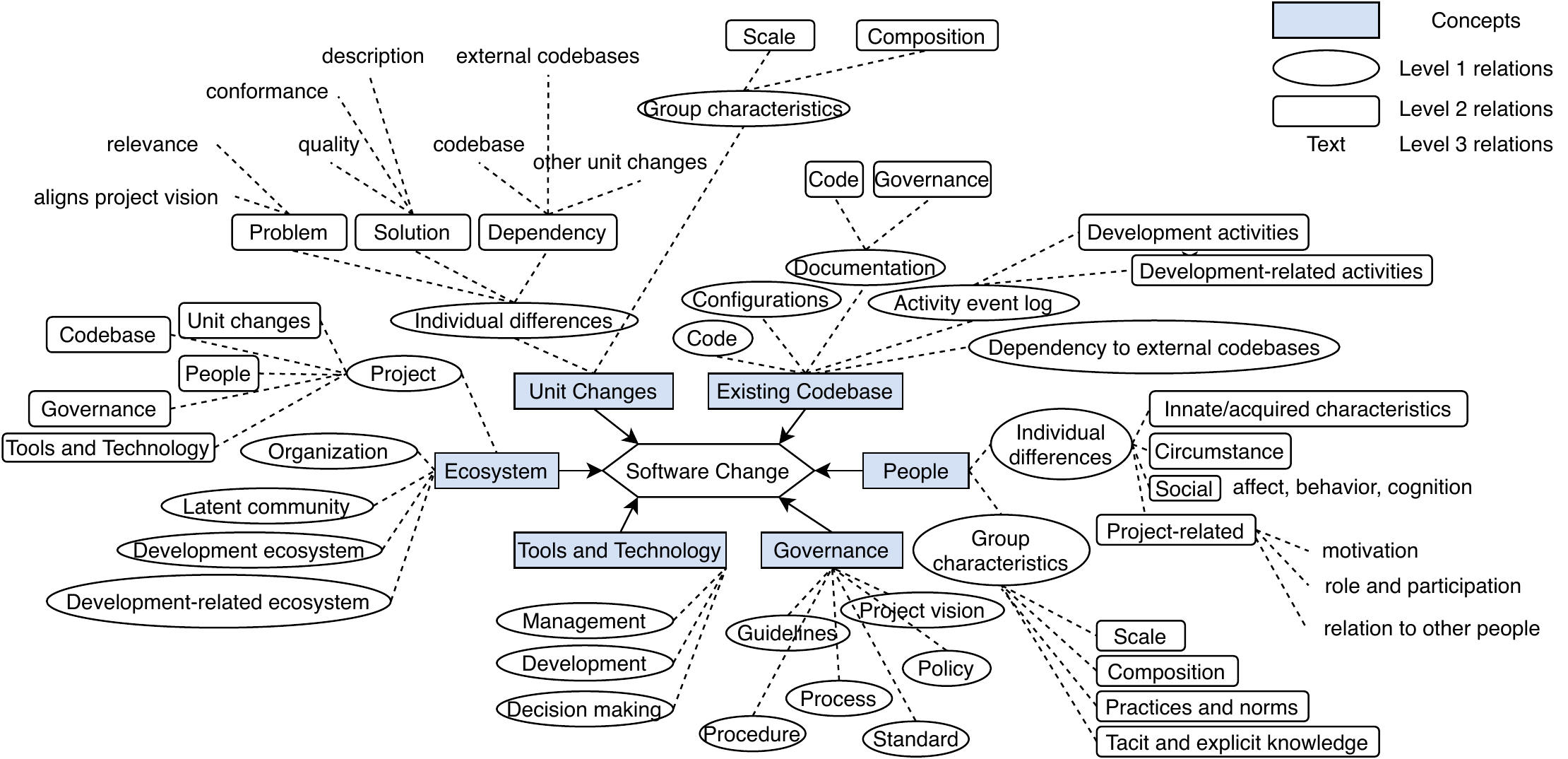}
\caption{A framework of concepts and categories influencing the decision of software change. 
The six core concepts influencing software change are shown in rectangles, also highlighted in blue.
The subsequent sub-categories of the six core concepts are arranged hierarchically and are shown in different shapes.}
\label{fig:initialTheoryFramework}
\end{figure*}

\section{Concepts and Initial Theory}
\label{sec:initialTheory}
% Overview
We identified six core concepts namely
\emph{unit changes, existing codebase, people, governance, tools and technology, and ecosystem}, influencing the decision of software change. 
In this section, we introduce categories and sub-categories of the six core concepts as a \emph{framework}
(see Figure~\ref{fig:initialTheoryFramework}).
% What does our initial theory looks like?
Using these high level concepts, we present our \emph{initial theory} depicting the mechanism of software change (see Figure~\ref{fig:initialTheory}).

\subsection*{Unit changes}
% Introduce how unit change relates to software change
Software changes by integrating \emph{unit changes} into an existing codebase. 
% Define unit change
A \emph{unit change} refers to a group of changes, generally code but also documentation and configurations, that is candidate for inclusion into an existing codebase.
Unit change is an umbrella term we use to refer to commonly used change approaches, such as pull request in GitHub\footnote{https://docs.github.com/en/github/collaborating-with-issues-and-pull-requests/about-pull-requests} and patches in Apache projects.\footnote{https://cocoon.apache.org/2.0/howto/howto-patch.html}
Unit change, however, is different from these terms since it disassociates the process of software change from the change itself.
For instance, patches are discussed via emails while pull requests via GitHub's interface; 
a unit change in itself is not tied to any review mechanism. 

% Explain how unit change influences software change in detail
Unit changes are proposed against the current state of a codebase. 
Multiple unit changes can be proposed in parallel.
Each unit change is uniquely described in terms of its individual characteristics as well as its relation to the other available unit changes (refer to the concepts and categories of \emph{unit changes} in Figure~\ref{fig:initialTheoryFramework}).
Individually, a unit change is characterized by the \emph{problem} it intends to solve and the proposed \emph{solution}.
The characteristics of a problem influencing the decision of software change relates to its \emph{alignment to project vision}~\cite{gousiosWorkPracticesChallenges2015b} but also relative \emph{relevance} (e.g., bug fix versus feature enhancement~\cite{padhyeStudyExternalCommunity2014}).
Likewise, a solution is characterized by its \emph{quality} (e.g., technical debt~\cite{pooputFindingImpactFactors2018}), \emph{description}~\cite{ramWhatMakesCode2018}, and \emph{conformance to project guidelines} (e.g., code style~\cite{ramWhatMakesCode2018}). 

A unit change is also characterized by its \emph{dependencies}, 
with some dependencies likely to cause conflicts (e.g., in case of large unit 
changes~\cite{diasUnderstandingPredictiveFactors2020}), influencing the decision of software change.
A unit change can have dependencies to \emph{other available unit changes, existing codebase}, as well the \emph{external codebases}.
Some of these dependencies are \emph{direct} (e.g., dependency to other unit changes~\cite{zhangHowMultiplePull2018a}) while others are \emph{indirect} (e.g., external codebase dependent on the changes to the existing codebase~\cite{blincoeReferenceCouplingExploration2019}). 
The nature of these dependencies can be \emph{syntactic} meaning that the unit change breaks other unit changes, existing codebase, and external codebases~\cite{cataldo2009software}.
Dependencies can also be \emph{semantic} meaning that while the code itself may not break, it changes the meaning or renders it meaningless~\cite{cataldo2009software}.
For instance, two unit changes solving the same problem renders one of the two redundant~\cite{liDetectingDuplicatePullrequests2017}.
Collectively, the \emph{scale} and \emph{composition} of available unit changes in a timeframe relates to workload~\cite{wangDuplicatePullRequest2019} and chances of conflict~\cite{liDetectingDuplicatePullrequests2017}, influencing the decision of software change.  

\subsection*{Existing Codebase}
% Describe the role of existing codebase in software change
Software changes by integrating unit changes into an \emph{existing codebase}. 
% Define existing codebase
In this process, the codebase transitions from its current state (referred to as \emph{existing codebase}) to a new state.
All subsequent unit changes are reviewed with respect to the new current state. 
Over time, the choice of unit changes integrated and discarded represents the trajectory of a software system evolution.

% Describe the different characteristics of a codebase influencing software change
Codebase has four characteristics: code, configurations, documentation, and activity event logs, 
in addition to its dependencies within and outside the codebase (refer to the concepts and categories of \emph{existing codebase} in Figure~\ref{fig:initialTheoryFramework}). 
Software system evolves by assimilating unit changes into one line of development or related lines of development (e.g., version), representing its \emph{configurations}. 
Code is also tied to \emph{documentation} useful for describing the code (e.g., code comments) as well as its governance (e.g., contributing guidelines and code of conduct\footnote{https://docs.github.com/en/github/building-a-strong-community/setting-up-your-project-for-healthy-contributions}).
Finally, \emph{logs} track the history of evolution by capturing \emph{development} (e.g., version control system~\cite{kalliamvakouIndepthStudyPromises2016}) and \emph{development-related} (e.g., issues~\cite{johnson2005improving}) activities and events.

Codebase has \emph{dependencies} within as well as to the external codebases.
Within, one part of a codebase can have dependency(-ies) to another part of the codebase, syntactically as well as semantically~\cite{cataldo2009software}.
As a result, changes to one part of the codebase can influence the code in the other part(s) of the codebase.
Likewise, a codebase has dependencies to external codebases on which the current codebase dependents on or other codebases
it provides functionality to~\cite{blincoeReferenceCouplingExploration2019}.
In all of the above cases, direct as well as indirect dependencies of a codebase influences the chance of a unit change translating into software change~\cite{diasUnderstandingPredictiveFactors2020}. 

\subsection*{Ecosystem}
% Describe the role of ecosystem in software change
Integration of unit changes into an existing codebase happens within an \emph{ecosystem} (refer to the concept \emph{ecosystem} in Figure~\ref{fig:initialTheoryFramework}). 
In an ecosystem, a \emph{project} is the most fundamental functional unit made of codebase, available unit changes, people in different roles, governance, and tools and technology. 
From hereon, the definition of an ecosystem broadens, based on its relations to the project (e.g., dependencies) and/or people in the project (e.g., developer community)~\cite{lungu2010small}.

Our recursive definition of software ecosystem starts with an \emph{organization} comprising of software projects developed within (e.g., Apache projects\footnote{https://projects.apache.org/}). 
\emph{Latent community} comprises of software projects related in one or more aspects (e.g., latent community of Haskell projects is linked by the programming language).
At even higher levels of abstraction, there are development and development-related ecosystems. 
\emph{Development ecosystems} comprise of related as well as unrelated software projects (e.g., GitHub development platform) while 
\emph{development-related ecosystem} refers to platforms which offers content or inspiration for software change~\cite{manesHowOftenWhat2019}. 
An example of development-related ecosystem is StackOverflow\footnote{https://softwareengineering.stackexchange.com/} where people, who may also be a part of development ecosystem~\cite{badashianInvolvementContributionInfluence2014}, discuss software engineering problems which can then be adopted as is or inspire unit changes. 
Another example of development-related ecosystem is Reddit\footnote{https://www.reddit.com/r/SoftwareEngineering/}(e.g.,~\cite{horawalavithanaMentionsSecurityVulnerabilities2019}).

\subsection*{Governance}
% Describe the role of Governance in software change
\emph{Governance} defines the mechanisms of software change in a project. 
% Describe the characteristics of governance influencing software change
Governance comprises of six parts: project vision, policy, standards, process, procedure, and guidelines. 
Irrespective of the style of governance (e.g., protective and equitable~\cite{alamiHowFOSSCommunities2020}), each project has \emph{policies} representing a system of principles or high-level expectations to guide decisions (e.g., non discrimination policy).
Policy is inspired by the \emph{project vision} (e.g., roadmap~\cite{gousiosWorkPracticesChallenges2015b}). 
Policies are translated into standards which has quantifiable requirements (e.g., Linux Standard Base\footnote{https://refspecs.linuxfoundation.org/lsb.shtml}).
Policy is also translated into \emph{process} describing the mechanism (e.g., code review) and \emph{procedure} with prescriptions (e.g., GitHub's interface\footnote{https://github.com/features/code-review/} for code review). 
Finally, guidelines (e.g., community guidelines\footnote{https://docs.github.com/en/github/site-policy/github-community-guidelines}) are recommendations open to interpretation.

\subsection*{People}
% Describe the role of people in software change
\emph{People} create unit changes and follow process and guidelines to decide on software change.
In this process, people introduce variability arising from their individual differences as well as characteristics as a group (refer to the concepts and categories from \emph{people} in Figure~\ref{fig:initialTheoryFramework}).

% Describe the characteristics of people influencing software change
Individuals are unique in their characteristics, circumstances, and relations to project(s) and other people.
These differences reflect in the process of decision making on software changes as well as decision.
Individuals have \emph{innate characteristics} (e.g., demographics~\cite{rastogiRelationshipGeographicalLocation2018b}), \emph{acquired characteristics} (e.g., experience and affiliation~\cite{kononenkoStudyingPullRequest2018a}), \emph{social aspects} (affect~\cite{destefanisMeasuringAffectsGithub2018}, behavior~\cite{yuExploringPatternsSocial2014}, and cognition~\cite{marlowImpressionFormationOnline2012}) and \emph{circumstances} (e.g., socio-economic status~\footnote{https://opensourcesurvey.org/2017/}) unique to them.
Adding to the variability, individuals adapt their behavior to their surrounding. 
\emph{Relating to project}, individuals are characterized by their \emph{motivation} to participate~\cite{dabbishSocialCodingGitHub2012a}
which further links to their \emph{role} (e.g., formal versus informal) and \emph{participation} (e.g., workload~\cite{wangWhyMyCode2019}). 
Also, social aspect of individuals reflects in their \emph{relation to other people} (e.g., trust~\cite{calefatoPreliminaryAnalysisEffects2017}).

Other than individual characteristics, characteristics of group: \emph{scale}, \emph{composition} (e.g., gender and tenure diversity~\cite{vasilescuGenderTenureDiversity2015b}), \emph{practices and norms} (e.g., legible pull request ~\cite{tsayInfluenceSocialTechnical2014b}), and \emph{knowledge} (e.g., tacit knowledge~\cite{elazharyNotSayContribution2019}) influence the process of software change.

\subsection*{Tools and Technology}
% Describe the role of tools in software change
While people are primarily responsible for deciding on whether to integrate a software change, 
their activities are mediated, facilitated, constraint, and sometimes even replaced by \emph{tools and technology}.
% Describe the characteristics of tools influencing software change
Tools and technology contribute to the \emph{management, development, and decision making} activities of people (refer to \emph{tools and technology} in Figure~\ref{fig:initialTheoryFramework}). 

Tools and technology manage \emph{development activities} (e.g., using milestone for tracking\footnote{https://docs.github.com/en/github/managing-your-work-on-github/tracking-the-progress-of-your-work-with-milestones}), \emph{team} (e.g., enables collaboration among remote contributors\footnote{https://docs.github.com/en/github/collaborating-with-issues-and-pull-requests}), \emph{project} (e.g., organize access and permissions based on role\footnote{https://docs.github.com/en/github/getting-started-with-github/access-permissions-on-github}), and \emph{ecosystem} (e.g., development platform GitHub). 
Tools facilitate development activities by performing 
\emph{repeated defined tasks} (e.g., identify outdated dependencies~\cite{mirhosseiniCanAutomatedPull2017}), 
\emph{intelligent tasks} (e.g., generate description for unit change~\cite{liuAutomaticGenerationPull2019}), 
\emph{offer compliance} (e.g., show test results~\cite{huUseBotsImprove2019a}), 
\emph{optimization} (e.g., reviewer recommendation~\cite{yangEmpiricalStudyReviewer2017}), 
\emph{scalability} (e.g., continuous integration tools~\cite{vasilescuQualityProductivityOutcomes2015a}),
and \emph{wider reach} (e.g., tools for specific platforms\footnote{https://github.com/mobile}). 
Tools also enable decision making by generating \emph{insights and 
awareness} (e.g., using visualization~\cite{oosterwaalVisualizingCodeCoverage2016}) for the improvement of individuals, teams, project, organization, and ecosystem.

Tools replace humans in performing clearly-defined, repeatable tasks.
Although a new generation of tools are edging to mimic human developers by learning from traces, also referred to as bots~\cite{lebeuf2017software}.
Generally, tools can reduce human error, time and effort, give feedback, optimize, simplify jobs, and provide insights.
Although adoption of tools can also raise issues: tools not working as expected, limit activities, and introduce other issues (e.g., bias in work distribution~\cite{pengExploringHowSoftware2019} and fatigue~\cite{mirhosseiniCanAutomatedPull2017}). 

% Describe initial theory
\subsection*{Initial Theory}
\begin{figure}
\centering
\includegraphics[width=0.48\textwidth]{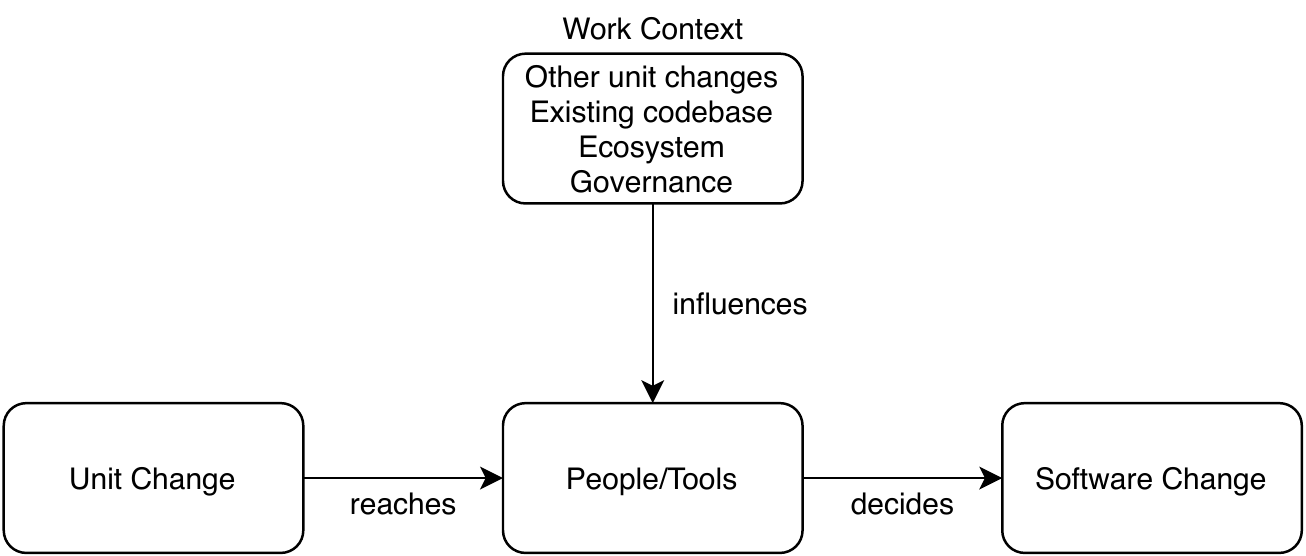}
\caption{Initial Theory of Software Change}
\label{fig:initialTheory}
\end{figure}

The six core concepts combined give a high-level view on the decision of software change (see Figure~\ref{fig:initialTheory}).
A unit change \emph{reaches} people and/or tools, 
who then \emph{decide} to integrated the unit change into existing codebase or discard it.
The decision of software change is influenced by the characteristics of \emph{unit change} but also the \emph{context} in which it exists.
Context for a unit change is created by \emph{governance} which defines standard process for software change,
including customizations provided by guidelines.
The dependencies of a unit change to the \emph{other unit changes} and \emph{existing codebase} within the \emph{project} and its broader \emph{ecosystem} also influences the decision of software change. 
Other than the above, characteristics of \emph{people}, and by extension \emph{tools}, involved in the process of software change influence the decision of software change. 

\begin{framed}
The \emph{decision of software change} is influenced by the characteristics of the \emph{unit change} and \emph{people/tool}, mediated by \emph{work context} (e.g., other unit changes and governance). 
\end{framed}

\section{Theory of Software Change}
\label{sec:finalTheory}
By adding relations to the concepts identified above, this section presents the mechanism of software change. 
For a given state of a software system, our theory explains the decision making mechanism preceding the software change. 

\begin{figure}
\centering
\includegraphics[width=0.47\textwidth]{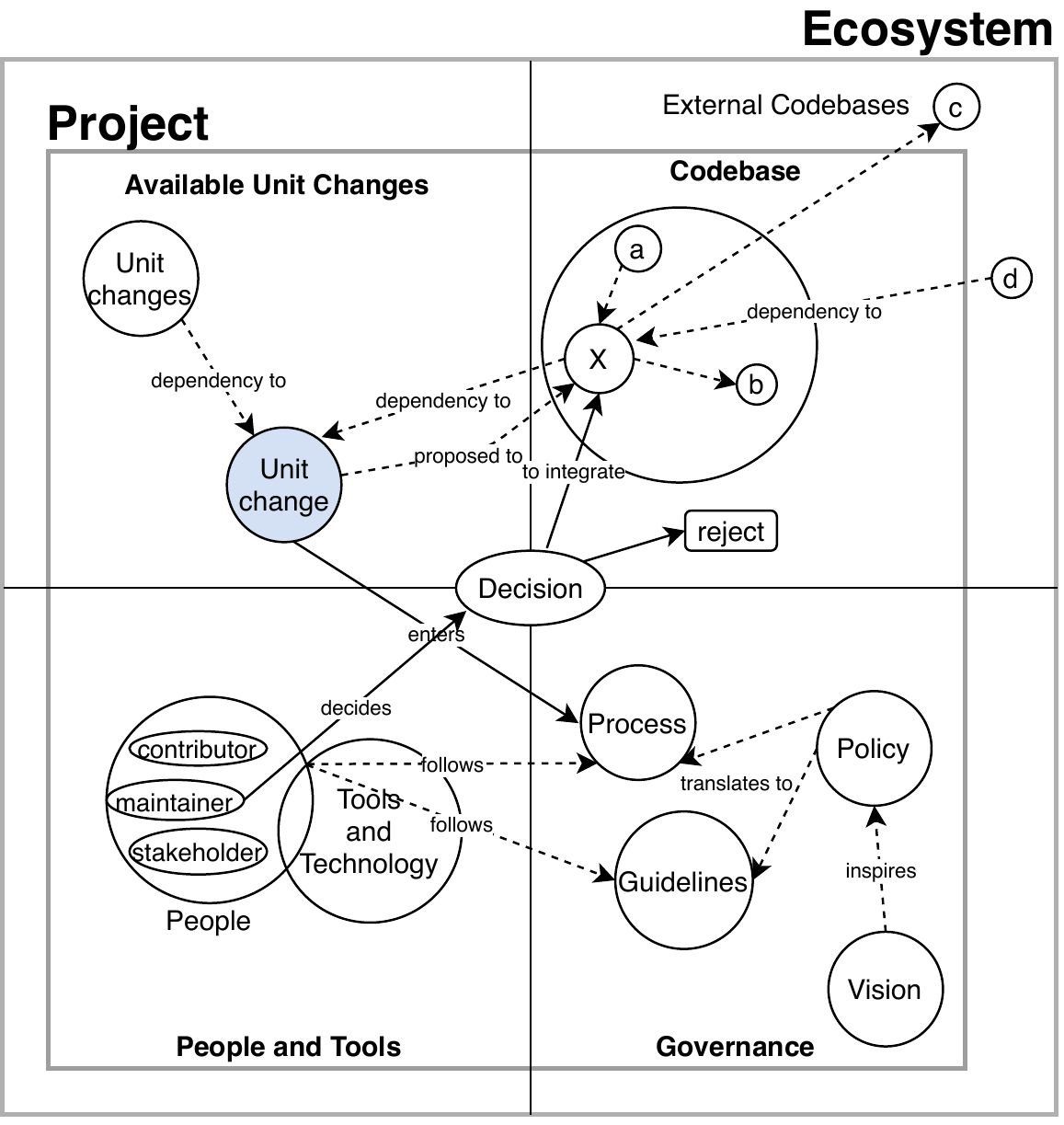}
\caption{Theory of Software Change}
\label{fig:finalTheory}
\end{figure}

We found that the decision of integrating a unit change is guided by four elements namely \emph{available unit changes, codebase, people and tools, and governance}. 
The four elements are a part of a project and its broader ecosystem, and are depicted as four quadrants in Figure~\ref{fig:finalTheory}.

The decision of software change, shown as an ellipse, is at the center of the four quadrants.
The process of software change is shown as a strong directional line connecting unit change to the decision and ultimately, the codebase.
All concepts are shown as circles connected by dotted directional lines indicating their relations. 

\emph{Available unit changes} (shown in second quadrant) are candidate for inclusion into an existing codebase.
At any given time, a unit change (shown in blue) is proposed against the current state of a part `X' of the existing codebase.
The decision to accept a unit change into existing codebase depends on its characteristics but also its dependency to other available unit changes.
For instance, unit change has syntactic and semantic dependencies to other unit changes causing competition (e.g., when two unit changes solve same problem~\cite{wangDuplicatePullRequest2019}) and/or conflict~\cite{diasUnderstandingPredictiveFactors2020}, 
influencing the decision of software change. 

A unit change has dependencies to the \emph{existing codebase} (shown in first quadrant) influencing the decision of software change. 
A unit change can have direct dependencies to the part of the codebase (X) where it is supposed to be merged (e.g., `a') and indirect dependencies otherwise (e.g., `d') coming from the external codebases that are part of the software ecosystem. 
Ultimately, the decision of software change is influenced by the characteristics of the unit change as well as its dependencies to other available unit changes and existing codebase.
Here, the existing codebase describes the context to which incoming unit change(s) should adapt or adopt.

The decision of software change is influenced by the process it undergo (see fourth quadrant).
The general process of software change is same for all incoming unit changes, as defined for a project.
The variability in the process arises based on project guidelines which are open and flexible. 
Eventually, the choice of process and guidelines followed by a project influences the decision of merging a unit change into existing codebase (e.g., CI tools increase the merge rate of unit changes~\cite{baltesNoInfluenceContinuous2018}). 
% Why no standard?

Finally, the characteristics of the \emph{people} who decides on the unit change influences the decision (see third quadrant).
As contributor, people create unit changes while as maintainer and stakeholder, people discuss the change with contributor and decide on the software change.
While people are expected to follow the process and guidelines defined by the governance to decide on software change,
the decisions are influenced by their individual characteristics as well as the characteristics of the group involved in decision making.
 
In the process of software change, people are supported by \emph{tools and technology}.
Tools can reduce the variability introduced by people in the decision of software change (e.g., reduce human error~\cite{mirhosseiniCanAutomatedPull2017}).
Although tools can also add variability as it learns from people (e.g., reinforcing bias~\cite{galhotra2017fairness}). 

In a nutshell, governance describes a standard process of software change with additional guidelines for customization.
When a unit change enters the process of software change, the decision to integrate is influenced by the characteristics of the unit change as well as its dependencies to other unit changes and existing codebase.
People follow the process and guidelines for software change but
introduce variability relating to individual differences as well as group characteristics.
Tools as extensions of people, on one hand, reduce variability introduced by people, but sometimes introduce variability of its own.
Ultimately, a unit change is either accepted for inclusion into the existing codebase or rejected.
In case the decision is an accept, unit change is merged, cherry-picked from, or an alternate solution to the same problem is proposed and integrated. 

With the integration of the unit change, codebase transitions from the existing state to a new current state.
All subsequent unit changes are reviewed with respect to the new current state of the codebase. 
The process of software change, as well as its decision, however, creates ripple effect (e.g., changes motivation of contributor~\cite{dabbishSocialCodingGitHub2012a}), discussion on which is beyond the scope of this theory.

%\begin{framed}
%Theory of software change has four core concepts: \emph{available unit changes} representing potential directions for software evolution; 
%\emph{codebase} and its dependencies describing the context; 
%\emph{governance} defining the process and guidelines for software change;
%and \emph{people and their extension tools} deciding on the software change.
%\end{framed}

\section{Discussion and Implications}
\label{sec:discussion}
In this section, we revisit our theory to describe its scope and theoretical quality.
Next, we present propositions derived from our theory. 
Finally, we discuss the implications of our theory on research, practice, and education. 

\subsection{Scope}
Since our theory is grounded in empirical data, 
first we describe the scope of our theory in terms of the evidence used for its formation - \emph{scope of validity}, followed by the scope of our theory, also referred to as \emph{scope of interest}.

\emph{Scope of Validity:} 
The evidence used for grounding our theory are based on open source as well as commercial projects hosted on GitHub.
Our theory builds on the insights drawn from thousands of software projects, with unique characteristics, community, policies, and more.
Also, we use a mix of studies reporting quantitative and qualitative insights.
Nonetheless, our theory has more evidence on open source software projects compared to company projects.
Also, the evidence incline more towards large and popular projects as against small projects.
Finally, our insights on privately developed software projects are limited. 

\emph{Scope of Interest:} 
Our theory is expected to explain the mechanism of software change given the current state of a software ecosystem until a decision is reached to integrate a unit change into existing codebase or discard it.
Once a decision is reached, we expect ripple effects to occur, 
changing not only the existing codebase but also its ecosystem. 
Describing the mechanism of ripple effects is beyond the scope of our theory.
Within the scope of our interest, we expect our theory to work on collaboratively developed modern software projects: small and large, open source and software companies, GitHub and outside.
% merge then review

% propositions and explanations
%!TEX root =theory.tex
\begin{table*}
\caption{Propositions and explanations derived from the theory of software change}
\label{tab:relationsTable}
\begin{tabularx}{\linewidth}{r X}
\hline
%\multicolumn{2}{l}{Constructs}\\ \hline
%C1 & Available unit changes\\
%C2 & Existing codebase\\
%C3 & People\\
%C4 & Governance \\
%C5 & Tools and technology\\
%C6 & Ecosystem\\
%\hline
\multicolumn{2}{l}{Propositions}\\ 
\hline
%P01 & Characteristics of a unit change influence the decision of software change.\\
\multicolumn{2}{l}{\emph{Investigated in the existing work}}\\
P01 & Unit changes compliant to the project vision are more likely to be integrated into existing codebase. \\
P02 & Unit changes with no dependency to other unit changes are more likely to be integrated into existing codebase. \\
P03 & Larger the impact of unit change; less likely it is to be integrated.\\
P04 & A well-modularized codebase increases the chance of unit change acceptance. \\ 
P05 & Decision to accept unit changes is affected by the signals that maintainers consciously or subconsciously consume through their interactions with contributors. \\
P06 & Tools and technology can improve the quality of a unit change, and therefore its chances of acceptance.\\
\\
\multicolumn{2}{l}{\emph{Emerging from the theory of software change}}\\

P07 & Unit changes that adapt to the technical norms of its ecosystem are more likely to be accepted.\\
P08 & Projects with strong tool support are more likely to scrutinize unit changes more.\\
P09 & Unit changes from new contributors that follow the guidelines are more likely to be accepted. \\
P10 & Unit changes from new contributors originating in the same ecosystem are more likely to be accepted. \\

\hline
\multicolumn{2}{l}{\emph{Explanations}}\\ 
\hline
E01 & Maintainers and stakeholders
% define the project vision and 
need to follow guidelines (e.g., project roadmaps that make project vision explicit) \\
E02 & Unit changes enter the review process; this entails cross-examination of the unit change along with other unit changes. Fewer cross-dependencies facilitate cross-examination. \\
E03 & Syntactic or semantic dependencies of a unit change to existing codebase and its transitive impact complicate the assessment of unit changes. \\
E04 & The impact of a unit change is easier to be assessed on well-architected projects, as their impact is contained. \\
E05 & The characteristics of people participating in the decision process influence the decision.\\
E06 & Automated feedback can improve both the internal quality and the adherence to project guidelines.\\
E07 & A project inherits standards and technical norms from its ecosystem (e.g. dependency declaration files); unit changes not using such standards may break the ecosystem.\\
E08 & Tools improve human abilities by automating repetitive tasks and enabling them to consume more information. Therefore, maintainers are better equipped to investigate unit changes in depth.\\
E09 & As trust between the contributor and the maintainer can not be established otherwise, adherence to guidelines is a signal for quality.\\
E10 & Trust is established by examining contributor profile within a set of interrelated projects. \\

\hline
%\multicolumn{2}{l}{Scope}\\ 
%\hline
% & Theory of software change explains evolution of a software system from a given state to the next. 
%On reaching the next stage, recursive effects on the elements of software change begin, which are not a part of this theory. 
%This theory is derived from and applies to collaboratively developed software projects developed in open source. 
%Although the theory is likely to be applicable in commercial software projects too.
%\\ \hline
\end{tabularx}
\end{table*}

\subsection{Theoretical Quality}
\emph{Testability:}
Our theory translates into propositions (see Table~\ref{tab:relationsTable}) that can be tested in other contexts for support.
Our theory can be empirically tested in case studies as well as surveys of collaboratively developed software projects.
We consider the testability of our theory as \emph{moderate}.

\emph{Empirical support:}
Our theory builds on a wide range of empirical evidence around software change.
We analyzed documentations, survey reports, secondary sources of information, selected software projects, and scientific papers presenting empirical evidence for our concepts and relations.
Since the proposed theory is largely backed by empirical evidence,
we consider the empirical support of this theory to be \emph{strong}.

\emph{Explanatory power:}
Many factors influence the decision of software change, all of which will be hard to measure.
We expect that the explanatory power of our theory is \emph{low}, as is also suggested for SE theories, in general~\cite{sjoberg2008building}.

\emph{Parsimony:}
We consider the parsimony of our theory as \emph{moderate} since 
we used a moderate number of constructs and propositions as compared to the factors found to influence the process of software change. 

\emph{Generality:} 
The generality of our theory is \emph{moderate} since the resulting theory is independent of the specific settings. 

\emph{Utility:}
This theory can be used with little adaptations by collaboratively developed software projects.
We consider the utility of our theory as \emph{high}.

\subsection{Propositions}
We derived ten propositions from our theory that are testable hypotheses to guide future research.
The complete list of propositions and their explanations derived from our theory is presented in Table~\ref{tab:relationsTable}. 
Some of the propositions derived from our theory are investigated in existing work, providing support to our theory.
For instance, our first proposition: \emph{``Unit changes compliant to the project vision are more likely to be integrated into existing codebase."}
 is supported by integrators' perspective to maintain vision of the project~\cite{gousiosWorkPracticesChallenges2015b} and factors influencing their review decision~\cite{alamiHowFOSSCommunities2020}. 
This proposition follows from our theory since maintainers and stakeholders need to follow guidelines (e.g., project roadmaps that make project vision explicit) for decision making.
We also proposed four propositions (P07 to P10) emerging from our theory but are not tested otherwise.

\subsection{Implications}
\label{sec:implications}
\emph{Research:}
Grounded in multiple sources of evidence in software engineering, 
our theory is a \emph{comprehension} of the existing body of knowledge.
This theory positions existing work in a \emph{frame of reference} while also \emph{guiding future research} on gaps.
Also, the propositions presented here are \emph{testable hypotheses} which the future research should verify in contexts beyond the scope of validity of this research.

\emph{Practice:}
Our theory can guide practitioners in
(a) optimizing the use of available resources (e.g., using tools for repetitive tasks), 
(b) minimize conflicts (e.g., by writing well-modularized code), and
(c) manage variability in the process of software change.
Generally, our theory can guide the discussions on tradeoffs (e.g., when it makes sense to have people do a task instead of tools and vice-versa?).

\emph{Education:}
Finally, our theory can help students as well as software engineers understand the \emph{big picture} on the decision of software change, in simpler terms.

\section{Related Work}
\label{sec:relatedWork}
We present seven selected theories in \se{} 
we took inspiration from, 
describes specific aspect of our theory,
 or addresses a closely related subject. 

\emph{Theory of coordination}~\cite{herbsleb2006collaboration} explains and helps predict ``how the distribution of decisions over developers and the density of constraints among decisions will affect development time, probability that a file contains a field defect, and developer productivity''. 
The theory models coordination of engineering decisions in software projects as a distributed constraint satisfaction problem.
It is a \emph{special case of our theory} modeling the relations of people and codebase dependencies to explain the process of software change (e.g., development time). 

\emph{Theory of motivation} explains the motivations of software engineers in terms of the job characteristics and its impact on job performance~\cite{franca2012towards}.
The paper presents a grounded theory generated from semi structured interviews conducted at many units (multi-case replication) in a small private software company. 
Our theory \emph{inherits} the characteristics of people relating to project from here.

\emph{Theory of distance} describes factors that can improve communication by impacting people, activities, and artefacts~\cite{bjarnason2016theory}. 
The paper presents a predictive theory that is generated inductively from interviews across companies and the initial hypotheses were inferred from previous research.
This theory \emph{relates} to the group characteristics of people in our theory.

\emph{Theory of software developer expertize} describes the factors influencing the formation of software developer expertize (in programming) over time~\cite{baltes2018towards}. 
The paper presents a teleological process theory grounded in the data from mixed-methods survey and literature on expertize and expert performance.
While our theory takes individual differences in people as given, this theory \emph{explains} differences in expertize.

\emph{Theory of job satisfaction and perceived productivity} presents relations between job satisfaction and perceived productivity and the factors that influence this relationship~\cite{storey2019towards}. 
The paper anticipates a bidirectional relation between job satisfaction and perceived productivity (as seen in Psychology) that is empirically investigated in \se{} via survey at a large software company and insights from literature. 

\emph{Theory of engagement} presents the process by which developers get engaged in open source software projects via programs such as Google Summer of Code~\cite{silva2020engagement}. 
The theory is grounded in multiple data sources: documentation, OSS projects, experiences of students and mentors, as well as literature.
The effectiveness of the theory is explained in terms of the perceptions of the non-participants on the assistance the theory offers to achieve their goal.

\emph{Theory of software design} describes creation of complex software systems in terms of three activities: sensemaking from the context, coevolution of the context and design space, and implementation of the understanding into a technical artifact~\cite{ralph2016software}.
The theory is grounded in literature within and outside software engineering. 
Similar to the theory of software design, our theory explains creation of software systems, albeit from the \emph{perspective of incremental unit changes}.
 
\section{Limitations and Threats to Validity}
\label{sec:threatsToValidity}
\par{\emph{Internal Validity}}: 
We collected evidence from multiple sources to present a comprehensive picture on the relations.
Nonetheless, it is possible that we did not find all the factors influencing software change.
Perhaps, with even more data, we could find different ways to categories concepts which could increase its explanatory power.

The classification of factors into concepts and categories has its limitations.
One, the categories for classification are not mutually exclusive.
For example, \emph{individual characteristics} and \emph{social aspects} refer to some overlapping factors. 
Nonetheless, we choose to keep both categories to emphasize its relevance.
Two, some factors can be hard to classify (e.g., Is gender an innate or acquired characteristic? -- referring to Figure~\ref{fig:initialTheoryFramework}).
Finally, there are many ways for classification and by choosing a classification we deem most relevant we are possibly missing on some information.

\par{\emph{External Validity:}}
Our theory builds on multiple sources of evidence relating to collaborative software development in GitHub.
This includes project developed in open source as well as software companies.
While the \emph{scope of validity} of our theory can raise questions on the \emph{scope of interest}, we believe that our theory is applicable to a wide range of collaboratively developed software projects.
Our confidence is grounded in the abstract nature of concepts and relations derived in our theory.

\par{\emph{Credibility:}}
Our theory is build inductively from the empirical evidence gathered from multiple data sources.
Our core concepts occurred frequently suggesting that we have reached theoretical saturation.
For under-represented categories in our empirical data, we have collected more evidence.  

\par{\emph{Originality:}}
To the best of our knowledge, we are the first to contribute a theory of software change in software engineering. 

\par{\emph{Usefulness:}}
Our theory is useful as a reference document, positioning existing and future research in a frame of reference.
Our theory also guides future research through propositions derived from our theory. 

\par{\emph{Other Limitations:}}
The qualitative analysis and theory-building was primarily conducted by the first author, followed by a discussion with the second author until they reached an agreement.
This process is likely to introduce bias which we tried to mitigate by grounding our observations in empirical data.
That being said, while building theory, there is always an ``uncodifiable step'' based on the imagination of researchers that can influence the result~\cite{weick1989theory}.

\section{Conclusion and Future Work}
\label{sec:conclusion}
Our theory surfaces six core concepts that must interact for software to change.
Project \emph{governance} describes the process and guidelines for software change. 
When a unit change enters the process of software change, the decision to integrate the change is influenced by the characteristics of the unit change as well as its dependencies to \emph{other unit changes}, \emph{existing codebase} and external codebases in the \emph{ecosystem}.
\emph{People} follow the process and guidelines for software change but
introduce variability relating to individual differences as well as group.
\emph{Tools} as extensions of people, further introduces variability of its own.
Our process theory is grounded in evidence from multiple sources including documentations, analysis of selected software projects, and scientific literature in software engineering. 

%\subsection*{What's next?}
There are many directions to pursue from here.
Our current work described the mechanism of software change for a given state of software system. 
In practice, software systems are dynamic with core entities co-evolving with the process of software change. 
The next edition of this theory should explain these dynamic relations.
 Our theory can also be seen as a first step towards building a variance theory that can explain in greater details when and how software changes.
Finally, it will be interesting to see application and adaptation of our theory to new situations. 
%\section*{Acknowledgment}
\bibliographystyle{IEEEtran}
\bibliography{bib/PR,bib/theory,bib/modernCodeReview,bib/systematicMapping,bib/bib,bib/misc}

\begin{thebibliography}{10}
\providecommand{\url}[1]{#1}
\csname url@samestyle\endcsname
\providecommand{\newblock}{\relax}
\providecommand{\bibinfo}[2]{#2}
\providecommand{\BIBentrySTDinterwordspacing}{\spaceskip=0pt\relax}
\providecommand{\BIBentryALTinterwordstretchfactor}{4}
\providecommand{\BIBentryALTinterwordspacing}{\spaceskip=\fontdimen2\font plus
\BIBentryALTinterwordstretchfactor\fontdimen3\font minus
  \fontdimen4\font\relax}
\providecommand{\BIBforeignlanguage}[2]{{%
\expandafter\ifx\csname l@#1\endcsname\relax
\typeout{** WARNING: IEEEtran.bst: No hyphenation pattern has been}%
\typeout{** loaded for the language `#1'. Using the pattern for}%
\typeout{** the default language instead.}%
\else
\language=\csname l@#1\endcsname
\fi
#2}}
\providecommand{\BIBdecl}{\relax}
\BIBdecl

\bibitem{lehman1979understanding}
M.~M. Lehman, ``On understanding laws, evolution, and conservation in the
  large-program life cycle,'' \emph{Journal of Systems and Software}, vol.~1,
  pp. 213--221, 1979.

\bibitem{rajlich2000staged}
V.~T. Rajlich and K.~H. Bennett, ``A staged model for the software life
  cycle,'' \emph{Computer}, vol.~33, no.~7, pp. 66--71, 2000.

\bibitem{bacchelli2013expectations}
A.~Bacchelli and C.~Bird, ``Expectations, outcomes, and challenges of modern
  code review,'' in \emph{2013 35th International Conference on Software
  Engineering (ICSE)}.\hskip 1em plus 0.5em minus 0.4em\relax IEEE, 2013, pp.
  712--721.

\bibitem{beller2014modern}
M.~Beller, A.~Bacchelli, A.~Zaidman, and E.~Juergens, ``Modern code reviews in
  open-source projects: Which problems do they fix?'' in \emph{Proceedings of
  the 11th working conference on mining software repositories}, 2014, pp.
  202--211.

\bibitem{gousiosExploratoryStudyPullbased2014b}
G.~Gousios, M.~Pinzger, and A.~van Deursen, ``An exploratory study of the
  pull-based software development model,'' in \emph{Proceedings of the 36th
  {{International Conference}} on {{Software Engineering}}}, ser. {{ICSE}}
  2014.\hskip 1em plus 0.5em minus 0.4em\relax {Hyderabad, India}: {Association
  for Computing Machinery}, May 2014, pp. 345--355.

\bibitem{sadowski2018modern}
C.~Sadowski, E.~S{\"o}derberg, L.~Church, M.~Sipko, and A.~Bacchelli, ``Modern
  code review: a case study at google,'' in \emph{Proceedings of the 40th
  International Conference on Software Engineering: Software Engineering in
  Practice}, 2018, pp. 181--190.

\bibitem{KRHDG19}
\BIBentryALTinterwordspacing
E.~Kula, A.~Rastogi, H.~Huijgens, A.~{van Deursen}, and G.~Gousios, ``Releasing
  fast and slow: An exploratory case study at ing,'' in \emph{Proceedings of
  the 27th ACM Joint European Software Engineering Conference and Symposium on
  the Foundations of Software Engineering}.\hskip 1em plus 0.5em minus
  0.4em\relax ACM DL, 2019. [Online]. Available:
  \url{/pub/releasing-fast-and-slow.pdf}
\BIBentrySTDinterwordspacing

\bibitem{yuDeterminantsPullbasedDevelopment2016a}
Y.~Yu, G.~Yin, T.~Wang, C.~Yang, and H.~Wang,
  ``\BIBforeignlanguage{en}{Determinants of pull-based development in the
  context of continuous integration},'' \emph{\BIBforeignlanguage{en}{Science
  China Information Sciences}}, vol.~59, no.~8, p. 080104, Jul. 2016.

\bibitem{gousiosDatasetPullbasedDevelopment2014c}
G.~Gousios and A.~Zaidman, ``A dataset for pull-based development research,''
  in \emph{Proceedings of the 11th {{Working Conference}} on {{Mining Software
  Repositories}}}, ser. {{MSR}} 2014.\hskip 1em plus 0.5em minus 0.4em\relax
  {Hyderabad, India}: {Association for Computing Machinery}, May 2014, pp.
  368--371.

\bibitem{ramWhatMakesCode2018}
A.~Ram, A.~A. Sawant, M.~Castelluccio, and A.~Bacchelli, ``What makes a code
  change easier to review: An empirical investigation on code change
  reviewability,'' in \emph{Proceedings of the 2018 26th {{ACM Joint Meeting}}
  on {{European Software Engineering Conference}} and {{Symposium}} on the
  {{Foundations}} of {{Software Engineering}}}, ser. {{ESEC}}/{{FSE}}
  2018.\hskip 1em plus 0.5em minus 0.4em\relax {New York, NY, USA}:
  {Association for Computing Machinery}, Oct. 2018, pp. 201--212.

\bibitem{soaresAcceptanceFactorsPull2015b}
D.~M. Soares, M.~L. {de Lima J{\'u}nior}, L.~Murta, and A.~Plastino,
  ``Acceptance factors of pull requests in open-source projects,'' in
  \emph{Proceedings of the 30th {{Annual ACM Symposium}} on {{Applied
  Computing}}}, ser. {{SAC}} '15.\hskip 1em plus 0.5em minus 0.4em\relax {New
  York, NY, USA}: {Association for Computing Machinery}, Apr. 2015, pp.
  1541--1546.

\bibitem{tsayInfluenceSocialTechnical2014b}
J.~Tsay, L.~Dabbish, and J.~Herbsleb, ``Influence of social and technical
  factors for evaluating contribution in {{GitHub}},'' in \emph{Proceedings of
  the 36th {{International Conference}} on {{Software Engineering}}}, ser.
  {{ICSE}} 2014.\hskip 1em plus 0.5em minus 0.4em\relax {New York, NY, USA}:
  {Association for Computing Machinery}, May 2014, pp. 356--366.

\bibitem{iyerEffectsPersonalityTraits2019}
R.~N. Iyer, S.~A. Yun, M.~Nagappan, and J.~Hoey, ``Effects of {{Personality
  Traits}} on {{Pull Request Acceptance}},'' \emph{IEEE Transactions on
  Software Engineering}, pp. 1--1, 2019.

\bibitem{wang2019evolution}
D.~Wang, Y.~Ueda, R.~G. Kula, T.~Ishio, and K.~Matsumoto, ``The evolution of
  code review research: A systematic mapping study,'' \emph{arXiv preprint
  arXiv:1911.08816}, 2019.

\bibitem{gousios2016work}
G.~Gousios, M.-A. Storey, and A.~Bacchelli, ``Work practices and challenges in
  pull-based development: the contributor's perspective,'' in \emph{2016
  IEEE/ACM 38th International Conference on Software Engineering (ICSE)}.\hskip
  1em plus 0.5em minus 0.4em\relax IEEE, 2016, pp. 285--296.

\bibitem{gregor2006nature}
S.~Gregor, ``The nature of theory in information systems,'' \emph{MIS
  quarterly}, pp. 611--642, 2006.

\bibitem{fagan1976design}
M.~Fagan, ``Design and code inspections to reduce errors in program
  development,'' \emph{IBM Systems Journal}, vol.~15, no.~3, pp. 182--211,
  1976.

\bibitem{rigby2013convergent}
P.~C. Rigby and C.~Bird, ``Convergent contemporary software peer review
  practices,'' in \emph{Proceedings of the 2013 9th Joint Meeting on
  Foundations of Software Engineering}, 2013, pp. 202--212.

\bibitem{ralph2018toward}
P.~Ralph, ``Toward methodological guidelines for process theories and
  taxonomies in software engineering,'' \emph{IEEE Transactions on Software
  Engineering}, vol.~45, no.~7, pp. 712--735, 2018.

\bibitem{sjoberg2008building}
D.~I. Sj{\o}berg, T.~Dyb{\aa}, B.~C. Anda, and J.~E. Hannay, ``Building
  theories in software engineering,'' in \emph{Guide to advanced empirical
  software engineering}.\hskip 1em plus 0.5em minus 0.4em\relax Springer, 2008,
  pp. 312--336.

\bibitem{shull2008building}
F.~Shull and R.~L. Feldmann, ``Building theories from multiple evidence
  sources,'' in \emph{Guide to Advanced Empirical Software Engineering}.\hskip
  1em plus 0.5em minus 0.4em\relax Springer, 2008, pp. 337--364.

\bibitem{charmaz2014constructing}
K.~Charmaz, \emph{Constructing grounded theory}.\hskip 1em plus 0.5em minus
  0.4em\relax sage, 2014.

\bibitem{herbsleb2006collaboration}
J.~Herbsleb and J.~Roberts, ``Collaboration in software engineering projects: A
  theory of coordination,'' \emph{ICIS 2006 Proceedings}, p.~38, 2006.

\bibitem{franca2012towards}
A.~C.~C. Franca, D.~E. Carneiro, and F.~Q. da~Silva, ``Towards an explanatory
  theory of motivation in software engineering: A qualitative case study of a
  small software company,'' in \emph{2012 26th Brazilian Symposium on Software
  Engineering}.\hskip 1em plus 0.5em minus 0.4em\relax IEEE, 2012, pp. 61--70.

\bibitem{bjarnason2016theory}
E.~Bjarnason, K.~Smolander, E.~Engstr{\"o}m, and P.~Runeson, ``A theory of
  distances in software engineering,'' \emph{Information and Software
  Technology}, vol.~70, pp. 204--219, 2016.

\bibitem{baltes2018towards}
S.~Baltes and S.~Diehl, ``Towards a theory of software development expertise,''
  in \emph{Proceedings of the 2018 26th acm joint meeting on european software
  engineering conference and symposium on the foundations of software
  engineering}, 2018, pp. 187--200.

\bibitem{storey2019towards}
M.-A. Storey, T.~Zimmermann, C.~Bird, J.~Czerwonka, B.~Murphy, and
  E.~Kalliamvakou, ``Towards a theory of software developer job satisfaction
  and perceived productivity,'' \emph{IEEE Transactions on Software
  Engineering}, 2019.

\bibitem{gousiosGHTorrentGitHubData2012}
G.~Gousios and D.~Spinellis, ``{{GHTorrent}}: {{GitHub}}'s data from a
  firehose,'' in \emph{Proceedings of the 9th {{IEEE Working Conference}} on
  {{Mining Software Repositories}}}, ser. {{MSR}} '12.\hskip 1em plus 0.5em
  minus 0.4em\relax {Zurich, Switzerland}: {IEEE Press}, Jun. 2012, pp. 12--21.

\bibitem{stol2016grounded}
K.-J. Stol, P.~Ralph, and B.~Fitzgerald, ``Grounded theory in software
  engineering research: a critical review and guidelines,'' in
  \emph{Proceedings of the 38th International Conference on Software
  Engineering}, 2016, pp. 120--131.

\bibitem{silva2020engagement}
D.~M. G. C. T. M.~G. Jefferson~Silva, Igor~Wiese and I.~Steinmacher, ``A theory
  of the engagement in open source projects via summer of code programs,''
  2020.

\bibitem{gousiosWorkPracticesChallenges2015b}
G.~Gousios, A.~Zaidman, M.-A. Storey, and A.~{van Deursen}, ``Work practices
  and challenges in pull-based development: The integrator's perspective,'' in
  \emph{Proceedings of the 37th {{International Conference}} on {{Software
  Engineering}} - {{Volume}} 1}, ser. {{ICSE}} '15.\hskip 1em plus 0.5em minus
  0.4em\relax {Florence, Italy}: {IEEE Press}, May 2015, pp. 358--368.

\bibitem{padhyeStudyExternalCommunity2014}
R.~Padhye, S.~Mani, and V.~S. Sinha, ``A study of external community
  contribution to open-source projects on {{GitHub}},'' in \emph{Proceedings of
  the 11th {{Working Conference}} on {{Mining Software Repositories}}}, ser.
  {{MSR}} 2014.\hskip 1em plus 0.5em minus 0.4em\relax {New York, NY, USA}:
  {Association for Computing Machinery}, May 2014, pp. 332--335.

\bibitem{pooputFindingImpactFactors2018}
P.~Pooput and P.~Muenchaisri, ``Finding {{Impact Factors}} for {{Rejection}} of
  {{Pull Requests}} on {{GitHub}},'' in \emph{Proceedings of the 2018 {{VII
  International Conference}} on {{Network}}, {{Communication}} and
  {{Computing}}}, ser. {{ICNCC}} 2018.\hskip 1em plus 0.5em minus 0.4em\relax
  {New York, NY, USA}: {Association for Computing Machinery}, Dec. 2018, pp.
  70--76.

\bibitem{diasUnderstandingPredictiveFactors2020}
K.~Dias, P.~Borba, and M.~Barreto, ``\BIBforeignlanguage{en}{Understanding
  predictive factors for merge conflicts},''
  \emph{\BIBforeignlanguage{en}{Information and Software Technology}}, vol.
  121, p. 106256, May 2020.

\bibitem{zhangHowMultiplePull2018a}
X.~Zhang, Y.~Chen, Y.~Gu, W.~Zou, X.~Xie, X.~Jia, and J.~Xuan, ``How do
  {{Multiple Pull Requests Change}} the {{Same Code}}: {{A Study}} of
  {{Competing Pull Requests}} in {{GitHub}},'' in \emph{2018 {{IEEE
  International Conference}} on {{Software Maintenance}} and {{Evolution}}
  ({{ICSME}})}, Sep. 2018, pp. 228--239.

\bibitem{blincoeReferenceCouplingExploration2019}
K.~Blincoe, F.~Harrison, N.~Kaur, and D.~Damian,
  ``\BIBforeignlanguage{en}{Reference {{Coupling}}: {{An}} exploration of
  inter-project technical dependencies and their characteristics within large
  software ecosystems},'' \emph{\BIBforeignlanguage{en}{Information and
  Software Technology}}, vol. 110, pp. 174--189, Jun. 2019.

\bibitem{cataldo2009software}
M.~Cataldo, A.~Mockus, J.~A. Roberts, and J.~D. Herbsleb, ``Software
  dependencies, work dependencies, and their impact on failures,'' \emph{IEEE
  Transactions on Software Engineering}, vol.~35, no.~6, pp. 864--878, 2009.

\bibitem{liDetectingDuplicatePullrequests2017}
Z.~Li, G.~Yin, Y.~Yu, T.~Wang, and H.~Wang, ``Detecting {{Duplicate
  Pull}}-requests in {{GitHub}},'' in \emph{Proceedings of the 9th
  {{Asia}}-{{Pacific Symposium}} on {{Internetware}}}, ser.
  Internetware'17.\hskip 1em plus 0.5em minus 0.4em\relax {New York, NY, USA}:
  {Association for Computing Machinery}, Sep. 2017, pp. 1--6.

\bibitem{wangDuplicatePullRequest2019}
Q.~Wang, B.~Xu, X.~Xia, T.~Wang, and S.~Li, ``Duplicate {{Pull Request
  Detection}}: {{When Time Matters}},'' in \emph{Proceedings of the 11th
  {{Asia}}-{{Pacific Symposium}} on {{Internetware}}}, ser. Internetware
  '19.\hskip 1em plus 0.5em minus 0.4em\relax {Fukuoka, Japan}: {Association
  for Computing Machinery}, Oct. 2019, pp. 1--10.

\bibitem{kalliamvakouIndepthStudyPromises2016}
E.~Kalliamvakou, G.~Gousios, K.~Blincoe, L.~Singer, D.~M. German, and
  D.~Damian, ``\BIBforeignlanguage{en}{An in-depth study of the promises and
  perils of mining {{GitHub}}},'' \emph{\BIBforeignlanguage{en}{Empirical
  Software Engineering}}, vol.~21, no.~5, pp. 2035--2071, Oct. 2016.

\bibitem{johnson2005improving}
P.~M. Johnson, H.~Kou, M.~Paulding, Q.~Zhang, A.~Kagawa, and T.~Yamashita,
  ``Improving software development management through software project
  telemetry,'' \emph{IEEE software}, vol.~22, no.~4, pp. 76--85, 2005.

\bibitem{lungu2010small}
M.~Lungu and M.~Lanza, ``The small project observatory: a tool for reverse
  engineering software ecosystems,'' in \emph{Proceedings of the 32nd ACM/IEEE
  International Conference on Software Engineering-Volume 2}, 2010, pp.
  289--292.

\bibitem{manesHowOftenWhat2019}
S.~S. Manes and O.~Baysal, ``How often and what {{StackOverflow}} posts do
  developers reference in their {{GitHub}} projects?'' in \emph{Proceedings of
  the 16th {{International Conference}} on {{Mining Software Repositories}}},
  ser. {{MSR}} '19.\hskip 1em plus 0.5em minus 0.4em\relax {Montreal, Quebec,
  Canada}: {IEEE Press}, May 2019, pp. 235--239.

\bibitem{badashianInvolvementContributionInfluence2014}
A.~S. Badashian, A.~Esteki, A.~Gholipour, A.~Hindle, and E.~Stroulia,
  ``Involvement, contribution and influence in {{GitHub}} and stack overflow,''
  in \emph{Proceedings of 24th {{Annual International Conference}} on
  {{Computer Science}} and {{Software Engineering}}}, ser. {{CASCON}}
  '14.\hskip 1em plus 0.5em minus 0.4em\relax {USA}: {IBM Corp.}, Nov. 2014,
  pp. 19--33.

\bibitem{horawalavithanaMentionsSecurityVulnerabilities2019}
S.~Horawalavithana, A.~Bhattacharjee, R.~Liu, N.~Choudhury, L.~O.~Hall, and
  A.~Iamnitchi, ``Mentions of {{Security Vulnerabilities}} on {{Reddit}},
  {{Twitter}} and {{GitHub}},'' in \emph{{{IEEE}}/{{WIC}}/{{ACM International
  Conference}} on {{Web Intelligence}}}, ser. {{WI}} '19.\hskip 1em plus 0.5em
  minus 0.4em\relax {New York, NY, USA}: {Association for Computing Machinery},
  Oct. 2019, pp. 200--207.

\bibitem{alamiHowFOSSCommunities2020}
A.~Alami, M.~L. Cohn, and A.~W{{a}}isowski, ``How {{Do FOSS Communities
  Decide}} to {{Accept Pull Requests}}?'' in \emph{Proceedings of the
  {{Evaluation}} and {{Assessment}} in {{Software Engineering}}}, ser. {{EASE}}
  '20.\hskip 1em plus 0.5em minus 0.4em\relax {New York, NY, USA}: {Association
  for Computing Machinery}, Apr. 2020, pp. 220--229.

\bibitem{rastogiRelationshipGeographicalLocation2018b}
A.~Rastogi, N.~Nagappan, G.~Gousios, and A.~{van der Hoek}, ``Relationship
  between geographical location and evaluation of developer contributions in
  github,'' in \emph{Proceedings of the 12th {{ACM}}/{{IEEE International
  Symposium}} on {{Empirical Software Engineering}} and {{Measurement}}}, ser.
  {{ESEM}} '18.\hskip 1em plus 0.5em minus 0.4em\relax {New York, NY, USA}:
  {Association for Computing Machinery}, Oct. 2018, pp. 1--8.

\bibitem{kononenkoStudyingPullRequest2018a}
O.~Kononenko, T.~Rose, O.~Baysal, M.~Godfrey, D.~Theisen, and B.~{de Water},
  ``Studying pull request merges: A case study of shopify's active merchant,''
  in \emph{Proceedings of the 40th {{International Conference}} on {{Software
  Engineering}}: {{Software Engineering}} in {{Practice}}}, ser.
  {{ICSE}}-{{SEIP}} '18.\hskip 1em plus 0.5em minus 0.4em\relax {Gothenburg,
  Sweden}: {Association for Computing Machinery}, May 2018, pp. 124--133.

\bibitem{destefanisMeasuringAffectsGithub2018}
G.~Destefanis, M.~Ortu, D.~Bowes, M.~Marchesi, and R.~Tonelli, ``On measuring
  affects of github issues' commenters,'' in \emph{Proceedings of the 3rd
  {{International Workshop}} on {{Emotion Awareness}} in {{Software
  Engineering}}}, ser. {{SEmotion}} '18.\hskip 1em plus 0.5em minus 0.4em\relax
  {New York, NY, USA}: {Association for Computing Machinery}, Jun. 2018, pp.
  14--19.

\bibitem{yuExploringPatternsSocial2014}
Y.~Yu, G.~Yin, H.~Wang, and T.~Wang, ``Exploring the patterns of social
  behavior in {{GitHub}},'' in \emph{Proceedings of the 1st {{International
  Workshop}} on {{Crowd}}-Based {{Software Development Methods}} and
  {{Technologies}}}, ser. {{CrowdSoft}} 2014.\hskip 1em plus 0.5em minus
  0.4em\relax {New York, NY, USA}: {Association for Computing Machinery}, Nov.
  2014, pp. 31--36.

\bibitem{marlowImpressionFormationOnline2012}
J.~Marlow, ``Impression formation in online collaborative communities,'' in
  \emph{Proceedings of the 17th {{ACM}} International Conference on
  {{Supporting}} Group Work}, ser. {{GROUP}} '12.\hskip 1em plus 0.5em minus
  0.4em\relax {New York, NY, USA}: {Association for Computing Machinery}, Oct.
  2012, pp. 293--294.

\bibitem{dabbishSocialCodingGitHub2012a}
L.~Dabbish, C.~Stuart, J.~Tsay, and J.~Herbsleb, ``Social coding in {{GitHub}}:
  Transparency and collaboration in an open software repository,'' in
  \emph{Proceedings of the {{ACM}} 2012 Conference on {{Computer Supported
  Cooperative Work}}}, ser. {{CSCW}} '12.\hskip 1em plus 0.5em minus
  0.4em\relax {New York, NY, USA}: {Association for Computing Machinery}, Feb.
  2012, pp. 1277--1286.

\bibitem{wangWhyMyCode2019}
Q.~Wang, X.~Xia, D.~Lo, and S.~Li, ``\BIBforeignlanguage{en}{Why is my code
  change abandoned?}'' \emph{\BIBforeignlanguage{en}{Information and Software
  Technology}}, vol. 110, pp. 108--120, Jun. 2019.

\bibitem{calefatoPreliminaryAnalysisEffects2017}
F.~Calefato, F.~Lanubile, and N.~Novielli, ``A preliminary analysis on the
  effects of propensity to trust in distributed software development,'' in
  \emph{Proceedings of the 12th {{International Conference}} on {{Global
  Software Engineering}}}, ser. {{ICGSE}} '17.\hskip 1em plus 0.5em minus
  0.4em\relax {Buenos Aires, Argentina}: {IEEE Press}, May 2017, pp. 56--60.

\bibitem{vasilescuGenderTenureDiversity2015b}
B.~Vasilescu, D.~Posnett, B.~Ray, M.~G. {van den Brand}, A.~Serebrenik,
  P.~Devanbu, and V.~Filkov, ``Gender and {{Tenure Diversity}} in {{GitHub
  Teams}},'' in \emph{Proceedings of the 33rd {{Annual ACM Conference}} on
  {{Human Factors}} in {{Computing Systems}}}, ser. {{CHI}} '15.\hskip 1em plus
  0.5em minus 0.4em\relax {New York, NY, USA}: {Association for Computing
  Machinery}, Apr. 2015, pp. 3789--3798.

\bibitem{elazharyNotSayContribution2019}
O.~Elazhary, M.-A. Storey, N.~Ernst, and A.~Zaidman, ``Do as {{I Do}}, {{Not}}
  as {{I Say}}: {{Do Contribution Guidelines Match}} the {{GitHub Contribution
  Process}}?'' in \emph{2019 {{IEEE International Conference}} on {{Software
  Maintenance}} and {{Evolution}} ({{ICSME}})}, Sep. 2019, pp. 286--290.

\bibitem{mirhosseiniCanAutomatedPull2017}
S.~Mirhosseini and C.~Parnin, ``Can automated pull requests encourage software
  developers to upgrade out-of-date dependencies?'' in \emph{2017 32nd
  {{IEEE}}/{{ACM International Conference}} on {{Automated Software
  Engineering}} ({{ASE}})}, Oct. 2017, pp. 84--94.

\bibitem{liuAutomaticGenerationPull2019}
Z.~Liu, X.~Xia, C.~Treude, D.~Lo, and S.~Li, ``Automatic generation of pull
  request descriptions,'' in \emph{Proceedings of the 34th {{IEEE}}/{{ACM
  International Conference}} on {{Automated Software Engineering}}}, ser.
  {{ASE}} '19.\hskip 1em plus 0.5em minus 0.4em\relax {San Diego, California}:
  {IEEE Press}, Nov. 2019, pp. 176--188.

\bibitem{huUseBotsImprove2019a}
Z.~Hu and E.~Gehringer, ``Use {{Bots}} to {{Improve GitHub Pull}}-{{Request
  Feedback}},'' in \emph{Proceedings of the 50th {{ACM Technical Symposium}} on
  {{Computer Science Education}}}, ser. {{SIGCSE}} '19.\hskip 1em plus 0.5em
  minus 0.4em\relax {Minneapolis, MN, USA}: {Association for Computing
  Machinery}, Feb. 2019, pp. 1262--1263.

\bibitem{yangEmpiricalStudyReviewer2017}
C.~Yang, X.~Zhang, L.~Zeng, Q.~Fan, G.~Yin, and H.~Wang, ``An {{Empirical
  Study}} of {{Reviewer Recommendation}} in {{Pull}}-based {{Development
  Model}},'' in \emph{Proceedings of the 9th {{Asia}}-{{Pacific Symposium}} on
  {{Internetware}}}, ser. Internetware'17.\hskip 1em plus 0.5em minus
  0.4em\relax {New York, NY, USA}: {Association for Computing Machinery}, Sep.
  2017, pp. 1--6.

\bibitem{vasilescuQualityProductivityOutcomes2015a}
B.~Vasilescu, Y.~Yu, H.~Wang, P.~Devanbu, and V.~Filkov, ``Quality and
  productivity outcomes relating to continuous integration in {{GitHub}},'' in
  \emph{Proceedings of the 2015 10th {{Joint Meeting}} on {{Foundations}} of
  {{Software Engineering}}}, ser. {{ESEC}}/{{FSE}} 2015.\hskip 1em plus 0.5em
  minus 0.4em\relax {New York, NY, USA}: {Association for Computing Machinery},
  Aug. 2015, pp. 805--816.

\bibitem{oosterwaalVisualizingCodeCoverage2016}
S.~Oosterwaal, A.~van Deursen, R.~Coelho, A.~A. Sawant, and A.~Bacchelli,
  ``Visualizing code and coverage changes for code review,'' in
  \emph{Proceedings of the 2016 24th {{ACM SIGSOFT International Symposium}} on
  {{Foundations}} of {{Software Engineering}}}, ser. {{FSE}} 2016.\hskip 1em
  plus 0.5em minus 0.4em\relax {New York, NY, USA}: {Association for Computing
  Machinery}, Nov. 2016, pp. 1038--1041.

\bibitem{lebeuf2017software}
C.~Lebeuf, M.-A. Storey, and A.~Zagalsky, ``Software bots,'' \emph{IEEE
  Software}, vol.~35, no.~1, pp. 18--23, 2017.

\bibitem{pengExploringHowSoftware2019}
Z.~Peng and X.~Ma, ``\BIBforeignlanguage{en}{Exploring how software developers
  work with mention bot in {{GitHub}}},'' \emph{\BIBforeignlanguage{en}{CCF
  Transactions on Pervasive Computing and Interaction}}, vol.~1, no.~3, pp.
  190--203, Nov. 2019.

\bibitem{baltesNoInfluenceContinuous2018}
S.~Baltes, J.~Knack, D.~Anastasiou, R.~Tymann, and S.~Diehl, ``({{No}})
  influence of continuous integration on the commit activity in {{GitHub}}
  projects,'' in \emph{Proceedings of the 4th {{ACM SIGSOFT International
  Workshop}} on {{Software Analytics}}}, ser. {{SWAN}} 2018.\hskip 1em plus
  0.5em minus 0.4em\relax {New York, NY, USA}: {Association for Computing
  Machinery}, Nov. 2018, pp. 1--7.

\bibitem{galhotra2017fairness}
S.~Galhotra, Y.~Brun, and A.~Meliou, ``Fairness testing: testing software for
  discrimination,'' in \emph{Proceedings of the 2017 11th Joint Meeting on
  Foundations of Software Engineering}, 2017, pp. 498--510.

\bibitem{ralph2016software}
P.~Ralph, ``Software engineering process theory: A multi-method comparison of
  sensemaking--coevolution--implementation theory and
  function--behavior--structure theory,'' \emph{Information and Software
  Technology}, vol.~70, pp. 232--250, 2016.

\bibitem{weick1989theory}
K.~E. Weick, ``Theory construction as disciplined imagination,'' \emph{Academy
  of management review}, vol.~14, no.~4, pp. 516--531, 1989.

\end{thebibliography}
\end{document}